\documentclass[twocolumn,showpacs,preprintnumbers,amsmath,amssymb,psfig]{revtex4}
\usepackage{graphicx}

\begin{document}
\title{Preferential Exchange: Strengthening Connections in Complex Networks.}
\author{G. Caldarelli$^{1,2}$, F. Coccetti$^{2}$, P. De Los Rios$^{3}$}
\affiliation{$^1$INFM UdR ROMA1 Dipartimento Fisica, Universit\`a di
Roma La Sapienza, Piazzale Aldo Moro 2 00185, Roma, Italy,}
\affiliation{$^2$Museo Storico della Fisica e Centro Studi e Ricerche Enrico Fermi,
Compendio Viminale, 00189 Roma, Italy}
\affiliation{$^3$Laboratoire de Biophysique Statistique, ITP-SB, Ecole Polytechnique F\'ed\'erale
de Lausanne, 1015 Lausanne, Switzerland.}
\date{\today}
\begin{abstract}
Many social, technological and biological interactions involve network relationships whose
outcome intimately depends on the structure of the network and on
the strengths of the connections. Yet, although much information
is now available concerning the structure of many networks, the
strengths are more difficult to measure. Here we show that, for
one particular social network, notably the e-mail network, a
suitable measure of the strength of the connections can be
available. We also propose a simple mechanism, based on
positive feedback and reciprocity, that can explain the observed
behavior and that hints toward specific dynamics of formation and
reinforcement of network connections. Network data from contexts different 
from social sciences indicate that power-law, and generally broad, distributions of 
the connection strength are ubiquitous, and the proposed mechanism has a wide range of
applicability. 
\end{abstract}
\maketitle

Networks are the most general framework to describe technological, biological,
social and other systems. The nodes of the network (Internet routers~\cite{PRVV01}, 
Web pages~\cite{HA99}, proteins~\cite{JMBO01},
species~\cite{SM01}, companies~\cite{GBCSC03} and so on) are 
linked by connections that are present or absent 
depending on the node relations we are interested in. In the case of the Internet and of the WWW
what is a connection is clear, being cables or hyper-links. In other cases connections
can depend on the definition: for example, we may say that proteins interact
if they physically stick to each other, or if one of the two promotes the
expression of the other. Species interact by predation in food-webs and in the case of companies 
one possible relation is given by the companies' portfolio. 
Social relations between individuals can be of many kinds and purposes, from
business~\cite{Granovetter74,Boorman75,Montgomery91} to mutual
assistance~\cite{BLM00} to friendships and others. The choice of the type of 
relation defines the network and its structure, but we need also the strength of 
the connections to fully characterize the network.
In the social context, for example, the strength of a relation is important
to determine which is the best route to pass information to
or gather information from somebody else in the system.  
Strong social ties may be regarded as preferential and reliable information channels.

All the above networks present the small-world property, {\it i.e.} the average distance 
between nodes grows only logarithmically with the size of the network.
As such, small-world networks are usually considered optimal to distribute or
collect information. Yet, whenever some of the connections become
unreliable, the effective average distance can become rather large~\cite{BBCHS03}.
Under this respect the reliability of a connection, and ultimately
the robustness of the network can be assessed by the strength of various connections.
The most recent studies indeed complement the attention to the network topology
with an investigation on the weights of edges~\cite{GBCSC03,BBPSV03}. 
Yet, although the weights of the connections are clearly very important,
their determination is a difficult task. Indeed it is relatively easy to 
decide whether two
individuals are connected or not (since the
existence of a link between them is essentially a binary variable). 
Instead it is much more difficult to quantify the strength of such a 
connection. How can we measure in an objective way how much two
people are, for example, friends to each other? Here we show that
for e-mail networks (a particular instance of social network) such a
measure is possible. We believe that this example
provides clues on the mechanism by
which the network connections form, develop and strengthen.
We also introduce a model, based on the idea of {\it preferential 
exchange}, whose applicability can in principle be extended
to other contexts. 

Modern computer networks are inherently social networks, 
since they link people and organizations and allow the exchange 
of information and communications~\cite{Wellman01}. 
In particular the exchange of
e-mails between people defines a paradigmatic example of
computer-supported social network that is the object of many
recent studies~\cite{Wellman01,EMB02,TWH03,EMS03,Johansen03}. In
e-mail networks a link between two people is established whenever
they exchange an e-mail (or a threshold number of
e-mails~\cite{TWH03}). By browsing the e-mail folders of an
individual (each folder represents a different e-mail sender), it
is easy to check that, after a few years, the number of
connections for the average person can grow to the hundreds. A
careful analysis of the network is therefore necessary to reveal
the presence of groups with common interests and purposes and the
hierarchical organization of these groups~\cite{TWH03,GDDGA02}.

\begin{figure}
\rotatebox{0}{\scalebox{0.35}{\includegraphics{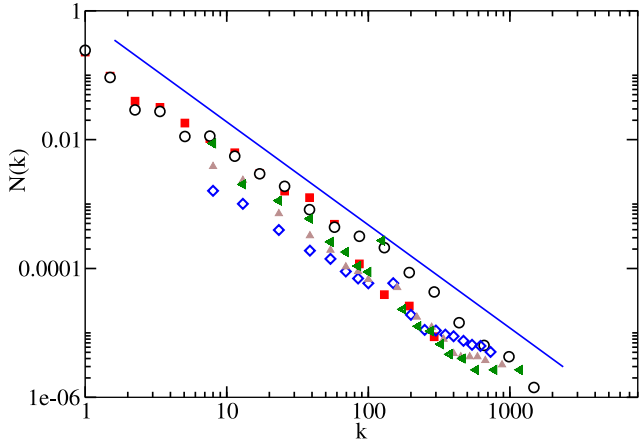}}}
\caption{Log-log representation of the e-mail distribution 
in five sets of folders (empty circles, full squares, and other symbols).
They are remarkably similar to each other, hinting towards some form of
universality. Data have been exponentially binned to reduce noise.
The straight line is a power-law $k^{-1.6}$.}
\label{Fig1}
\end{figure}

We introduce an objective measure of the strength of
the relations by keeping track of the number of e-mails received
from a given sender in somebody's e-mail directory. 
The data sets that we analyze are five e-mail directories coming from
our accounts and the accounts of two other colleagues.
They contain 
$5628$ e-mails (corresponding to $393$ senders) collected over
roughly three years, $19219$ e-mails ($476$ senders, ten years), 
$16102$ e-mails ($113$ senders, three years), $13385$ e-mails ($516$ senders, five years) 
and $21782$ e-mails ($207$ senders, five years).
Fig.\ref{Fig1} shows the normalized
histograms of the number $N(k)$ of people who wrote $k$ e-mails to us 
and our colleagues.
As it can be seen, they are quite similar,
and they can be approximated by an algebraic behavior of the kind
$N(k) \sim k^{-\gamma}$ with $\gamma \sim 1.6$. 
Although,
of course, the five datasets contain some common acquaintances,
they are mostly uncorrelated, so that we consider them to be
well representative of the same universal behavior.

An algebraic law, rather than a simple exponential, is usually a
symptom of the presence of some form of correlations in the
dynamical process that produced the data. How do correlations
arise in this context? A very simple mechanism that is known to
produce such correlations is a form of positive feedback that, in
the social context, can be described as "good partners become
better partners". Stated otherwise, there is a
reinforcement mechanism such that if the relation between two
people is already strong, it has more chances to become even
stronger.

To check whether this mechanism allows for the creation and
reinforcement of social links in such a way to reproduce the
empirical data, we have analyzed a very simple model. 
Starting from a society of $S_0$ individuals, at every time-step each of
them sends to the others $M_{out}$ e-mail messages, at random. 
The network of acquaintances grows in time, and at every time-step a 
new individual enters the society. 
The probability that individual $j$ sends a message to
individual $i$ is proportional to the number $k(i \to j)$ of
e-mails that $j$ ever received from $i$, that is
\begin{equation}
p(j \to i) = \frac{k(i \to j)}{\sum_l k(l \to j)}
\label{probability}
\end{equation}
(the sum in the denominator is the total number of e-mails ever
received by $j$). We assigned to this rule the name of {\it preferential
exchange}. 
In some respect this choice is reminiscent of the idea
of preferential attachment in the formation of growing scale-free 
networks~\cite{BA99}, even if, as we discuss in the following,
the physical meaning is rather different. More generally, the preferential exchange
is also close in spirit to the Tit-for-Tat
reciprocity strategy believed to be an important ingredient to explain the
emergence of cooperation and altruism between individuals~\cite{Axelrod}. 

\begin{figure}
\rotatebox{0}{\scalebox{0.35}{\includegraphics{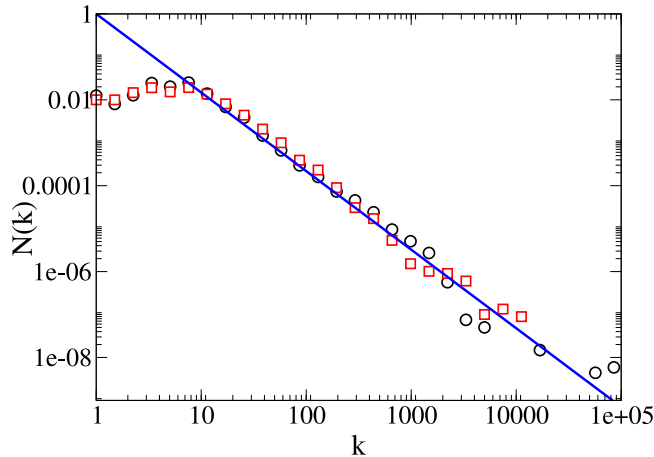}}}
\caption{Log-log representation of the e-mail distribution of two
random individuals of the model, with $S_0=2$, $S=2000$ and
$M_{out}=100$. The best power-law fit yields
an exponent $1.8(2)$ (straight line).} 
\label{Fig2}
\end{figure}

Fig.\ref{Fig2} shows the results of simulations with $S_0=2$,
followed for $1998$ time-steps, to a final size of $S=2000$ individuals; at 
every time-step each individual sends out $M_{out}=100$ e-mails.
As a starting condition we assume that every new individual has already 
exchanged one e-mail with everybody else: thus, the structure of the e-mail 
network is trivial, being fully connected. 
The e-mail distributions of random individuals in the population are
very similar to each other and all exhibit the same algebraic
behavior $N(k) \sim k^{-\gamma}$ with an exponent $\gamma \sim 1.8(2)$.
Noticeably the result does not depend on the choice of the above parameters.

The solution of the model can be obtained also analytically, by
means of a few approximations that allow for the identification of
the parameters relevant for the model. Indeed, it is possible to
write a rate equation for $k(j \to i, t)$:
\begin{equation}
\frac{dk(i \to j)}{dt} = M_{out} \cdot p(j \to i) = M_{out} \frac{k(i\to
j)}{\sum_l k(l \to j)} \label{rate equation}
\end{equation}
We assume that an individual receives e-mails at a constant rate
$M_{in}$, so that the denominator in the \textit{r.h.s.} of
(\ref{rate equation}) grows linearly in time: $M_{in} \cdot t$. We 
have verified this linear dependence on time in our simulations, 
finding furthermore that $M_{in} \simeq M_{out}$. 
Moreover, we assume that there is reciprocity in
the e-mail exchange, that is, the number of e-mails that $i$ ever
sent to $j$ is proportional to the number of e-mails that $j$ ever
sent to $i$. This allows us to replace the numerator
of the {\it r.h.s.} of (\ref{rate equation}) using $k(i \to j) = R \cdot k(j \to i)$. 
We have verified also this proportionality
in our simulations, finding $R\simeq 1$, an indication of the so-called 
{\it fair reciprocity}.  
With these assumptions the rate equation simplifies to
\begin{equation}
\frac{dk(j \to i,t)}{dt} = \alpha \frac{k(j \to i,t)}{t}
\label{simpler}
\end{equation}
with $\alpha = R(M_{out}/M_{in})$. 
The solution of Eq.(\ref{simpler}) is
\begin{equation}
k(j \to i,t)= \left(\frac{t}{t_0}\right)^{\alpha}.
\label{solution}
\end{equation}

If $t_i$ ($t_j$) is the time at which individual $i$ ($j$) entered the society, 
we set $t_0=max(t_i,t_j)$ (and of course $t_0 < t$). If $j$ is younger than $i$ then
$t_0 = t_j$ and we can invert the solution (\ref{solution}) to obtain
\begin{equation}
t_j = t [k(j \to i)]^{-\frac{1}{\alpha}}.
\label{inversion}
\end{equation}
Eq.(\ref{inversion}) sets a one-to-one relation between $t_j$ and $k(j \to i)$
that allows to use the probability conservation relation $N(k)dk = \rho(t) dt$,
where $\rho(t) = const$ because 
new individuals are added at a constant rate. Therefore we have that
$N(k) \sim k^{-\gamma}$ with $\gamma = 1+( M_{in}/M_{out})/R$. 
If on the contrary $j$ is older than
$i$, then $t_0 = t_i$ and these folders should contribute to a
peak of $N(k)$ at $k=(t/t_i)^{\alpha}$ independent of $j$. We do
not observe this peak in our simulations: if we split the
histogram of individual $i$ into the two contributions of people
older and younger than $i$, we find that they show the same
algebraic behavior (data not shown). This is due to the mean-field nature of
the above calculations. Fluctuations therefore
have been neglected.
This does not apply in the real situation where they are enhanced 
by the positive feedback mechanism. As a consequence their combined effect
drives the system to the same distribution $N(k)$ for individuals
both younger and older than $i$.
In the case of perfect reciprocity ($R=1$) and if people reply to every
e-mail they receive ($M_{in}/M_{out} = 1$), then the value of the exponent
$\gamma=2$, close to the results from our simulations. 

Actually, some of the approximations that we made can be safely
relaxed. In particular we might assume that, depending on the
personality, some people have a tendency to write slightly more
e-mails than they receive, {\it i.e.} $M_{out}/M_{in} > 1$, or vice-versa
(although very large or very small values are unreasonable and we
expect real values to be close to $1$); also, reciprocity could be
imperfect, always for personality reasons, and $R \neq 1$ (but
again very large or small values are unreasonable; this has been
again verified in our simulations). In these cases we can expect
variations of the exponent $\gamma$, (although nothing forbids
large variations of this exponent, our expectations are that
the exponents should always be close to $2$, as the data in Fig.\ref{Fig1} show).
Changing the values of $S$, $S_0$, and $M_{out}$ does not
change the results in our simulation. 

Our model, based on the {\it preferential exchange} ingredient, 
reproduces rather nicely the behavior of the data for a large range of 
parameter values.
As previously observed, this mechanism is similar to the preferential attachment model
proposed by Barab\'asi and Albert~\cite{BA99} to explain the emergence of the scale-free topology of
some networks. The mathematical similarity extends also to some other results: if, for example, 
the preferential exchange rate equation is modified so that the numerator in the {\it r.h.s.} of
(\ref{rate equation}) becomes $k(i\to j)^{\alpha}$, 
then the e-mail distribution becomes a stretched exponential, 
as it happens in the context of network topology~\cite{KRL00}.

Nevertheless relevant differences between the two rules appear when considering 
the nature of social networks.
Firstly, preferential exchange works on a local basis, which means that
two people can increase the strength of their link ignoring what 
is happening to the other links. 
Instead, in the preferential attachment model the newcomers need
a full knowledge of the network degrees in order to decide their connectivity.  
Secondly, and more importantly, the rate of change
of the e-mails that individual $i$ receives from $j$ depends only on the number of e-mails
that traveled in the opposite direction and on the total
number of messages that $j$ ever received (both local quantities available
to the two people).
Therefore preferential exchange is intrinsically {\em symmetric}, while preferential 
attachment divides the topology of the network in hubs and poorly connected nodes.
In summary, this is a symmetrically cooperative model where no global information 
is necessary.

Interestingly, more data have recently emerged about the connection strengths in scientific 
collaborations networks~\cite{LC03}, airport traffic~\cite{BBPSV03} 
and other systems, showing that the measured
strengths are indeed power-law, or at least-fat tail, distributed.
Networks often evolve through relations that get stronger in time
thanks to positive feedback, that is, the more an individual (in the social context)
has given to another one, the more the latter is likely to give back
in return. Moreover, many networks also grow in time.
Implementing these ingredients
in a simple model nicely reproduces the qualitative (algebraic)
behavior that we observe in real e-mail data. The quantitative
agreement is obtained when we add good reciprocity: the
exchange is a "fair" process. We believe that these ingredients do
indeed shape social and other networks, and the e-mail network, as a
particular example, is extremely suited to provide us with a
wealth of data that could be difficult to gather for other
networks (it has been found recently that in a sample of mailboxes
at the HP Laboratories the median number of e-mails was $2200$
indicating that a large amount of data could be, in principle,
available for analysis~\cite{WHAT03}). Moreover, we still
neglected the interplay between the dynamics over the network, and
the network structure itself, whereas we expect, in principle,
that the two should co-evolve toward some stationary state.

At the same time we expect that in most real situations this model
could be refined by introducing a more detailed description of the 
process of interaction.
For example, a large variability in people attitude could be captured  
by defining a local intrinsic quantity shaping the mechanism of link 
reinforcement.
As in the case of the preferential attachment mechanism such 
generalization does not remove the power-law nature of the probability distributions 
involved~\cite{BB01,GKK01,CCDM02,BM03}, but rather qualifies the kind of critical processes 
going on in the system. 
Further work is needed in this direction, and more data about the
structure of networks and the strength of the connections should be made 
available to develop and validate models.

From a more general point of view, e-mail networks on one hand, and
simulations on the other, can help investigate the large scale
consequences of fairness and reciprocity: these two ingredients
are often deemed as determinant in shaping social relations, yet
their effects are usually studied for small groups of people and
short times. The use of computers, both as data resources and as
simulation tools, can easily bring these studies to large scales.

We thank the FET-Open project IST-2001-33555 COSIN for
financial support. P.D.L.R. thanks the OFES-Bern for financial
support under contract 02.0234. 
We also thank Antonio Lanza and Furio Ercolessi 
for providing their e-mail folders. We acknowledge enlighting discussions
with M. Buchanan and R. Pastor-Satorras.

\end{document}